# High Curie temperatures at low compensation in the ferromagnetic semiconductor (Ga,Mn)As

M. Wang[1], K. W. Edmonds[1], B. L. Gallagher[1], A. W. Rushforth[1], O. Makarovsky[1], A. Patanè[1], R. P. Campion[1], C. T. Foxon[1], V. Novak[2], T. Jungwirth[2,1]

[1] *School of Physics and Astronomy, University of Nottingham, Nottingham NG7 2RD, United Kingdom*

[2] *Institute of Physics ASCR, v.v.i., Cukrovarnická 10, 16253 Praha 6, Czech Republic*

We investigate the relationship between the Curie temperature $T_C$ and the carrier density $p$ in the ferromagnetic semiconductor (Ga,Mn)As. Carrier densities are extracted from analysis of the Hall resistance at low temperatures and high magnetic fields. Results are found to be consistent with ion channeling measurements when performed on the same samples. We find that both $T_C$ and the electrical conductivity increase monotonically with increasing $p$, and take their largest values when $p$ is comparable to the concentration of substitutional Mn acceptors. This is inconsistent with models in which the Fermi level is located within a narrow isolated impurity band.

The III-V semiconductor (Ga,Mn)As is one of the most widely studied diluted magnetic materials exhibiting carrier-mediated ferromagnetism. Such systems are characterized by strong coupling between spin, charge and lattice degrees of freedom that may be modified by, for example, piezoelectrically applied stress or local electric fields [1]. (Ga,Mn)As also shows large anomalous magnetotransport effects that provide a 'built-in' sensor of the local magnetization orientation [1,2,3]. In particular, the anomalous Hall resistance ($R_{xy}^{AH}$) is proportional to the perpendicular-to-plane component of the magnetization, and offers a sensitive means of probing magnetization and domain wall dynamics in (Ga,Mn)As thin films [3]. On the other hand, the large magnitude of $R_{xy}^{AH}$ has hampered the determination of the carrier density, $p$, which is a crucial parameter for establishing a theoretical description of (Ga,Mn)As. $R_{xy}^{AH}$ is typically 10-100 times larger than the ordinary Hall resistance, $R_{xy}^{OH}$, and depends in a non-trivial way on various scattering mechanisms which contribute to the resistivity $\rho$. Hence, $R_{xy}^{AH}$ can dominate the slope of the Hall resistance even at high magnetic fields and at temperatures well above the Curie temperature, $T_C$.

Previous studies have utilized high magnetic fields and low temperatures in order to saturate the magnetization in the perpendicular-to-plane direction, and then separated the $R_{xy}^{OH}$ and $R_{xy}^{AH}$ contributions to the residual slope by fitting, in order to determine the carrier density in (Ga,Mn)As [4,5,6,7,8]. These measurements yielded close to one hole per substitutional Mn in layers with the highest $T_C$ values, which were annealed in air at temperatures comparable to or lower than the growth temperature [6]. Low-temperature annealing is a well-established technique to promote the out-diffusion of Mn interstitial ($Mn_I$) defects, which compensate holes provided by the ferromagnetic substitutional Mn ($Mn_S$) in (Ga,Mn)As [9,10].

Recently, ion channeling methods have been used to indirectly determine the carrier density in (Ga,Mn)As, by measurement of the concentrations of single acceptor $Mn_S$ and double donor $Mn_I$ [11]. The underlying assumption is that compensation by other types of defects (*eg* As antisites and Ga vacancies) is negligible, which may be questionable under some growth conditions [12,13,14]. On the other hand, this method has the advantage of avoiding the uncertainties associated with the magnetic field dependent anomalous Hall effect. The results obtained in Ref. [11] indicated a strong suppression of $T_C$ in (Ga,Mn)As layers with close to one hole per substitutional Mn, in striking disagreement with the earlier high field Hall effect measurements [6,7]. It was thus suggested that $T_C$ is determined by the location of the Fermi level within a narrow impurity band, separated from the valence band [11]. A similar model has been used to explain anomalous behaviors of the infrared conductivity and resonant tunneling spectroscopy in (Ga,Mn)As structures [15,16].

The aim of the present paper is to investigate the extent to which high-field Hall effect measurements can reliably be used to obtain the carrier density in (Ga,Mn)As, and thus to shed light on the discrepancy between Hall and ion channeling measurements of *p*. We present measurements on a wide range of (Ga,Mn)As thin film samples, including one in which *p* is varied in a series of short annealing steps. We also compare the value of *p* obtained from Hall measurements to previously published [17] ion channeling results for the same samples. Our results show no indication of a suppression of $T_C$ when $p$ approaches the concentration of substitutional Mn.

A range of (Ga,Mn)As films of thickness 25nm or 50nm were grown by molecular beam epitaxy. The total Mn concentration $x_{total}$, estimated from the Mn/Ga flux ratio during growth, varied from 1.5% to 12%. With increasing Mn flux, the growth temperature was lowered in order to maintain a two-dimensional growth mode, as described in detail elsewhere [18,19,20]. One of the films, with $x_{total}$=12%, was co-doped with hydrogen in order to increase the compensation in the as-grown state. H in (Ga,Mn)As is known to lead to a reduction of $p$, due to either compensation of substitutional Mn acceptors by interstitial H donors, or the formation of Mn-H complexes [21]. Furthermore, the H ions are weakly bound and out-diffuse on annealing at temperatures below 200°C [22]. Therefore, the film was annealed in air at temperatures from 120°C to 180°C, for small time steps in order to vary the magnetic and electrical properties over a wide range. The temperature-dependence of the resistivity for as-grown and annealed $x_{total}$=12% samples, with and without H co-doping, is plotted in Fig. 1, showing similar characteristics to previously studied samples which are doped above the metal-insulator transition [23].

The longitudinal and Hall resistances, $R_{xx}$ and $R_{xy}$, were recorded simultaneously versus magnetic field for Hall bar samples fabricated from the (Ga,Mn)As thin films. Representative measurements, for the annealed H co-doped sample, are shown in Fig. 2. At low fields the resistances vary as the magnetization is reoriented from the easy-plane to the hard out-of-plane direction by the applied magnetic field. Our focus is on the higher field parts of the resistance traces. The $R_{xy}$ is usually described as the sum of ordinary and anomalous parts:

$R_{xy} = R_0 B_z + R_A \mu_0 M_z$ (1)

where $B_z$ and $M_z$ are the perpendicular-to-plane components of the magnetic field and the magnetization. The ordinary Hall coefficient $R_0$ is given by ($r_H / ped$), where $r_H$ is the Hall factor, $p$ is the hole density, $e$ the electronic charge and $d$ the film thickness. Numerical calculations incorporating exchange and spin-orbit splitting have shown that $r_H$=1 to within an accuracy of 20% for highly $p$-doped (Ga,Mn)As [7]. The anomalous Hall coefficient $R_A$ is proportional to $R_{xx}^n$, where $n$ depends on the mechanism giving rise to the anomalous Hall resistance [24,25]. Experimentally, it is typically found that $n \approx 2$ in highly doped, lightly compensated (Ga,Mn)As films [5,8,26], while $n \leq 1$ has been reported for highly compensated samples [8,26].

The $R_{xy}$ versus $B$ curves were fitted assuming that the magnetization $M_z$ reaches a constant, saturated value above 3T, so that eqn. (1) reduces to

$R_{xy} = R_0 B_z + C R_{xx}^n$,

where $C$ is a fitting parameter, and $n = 1$ or 2. For the fully annealed films, the magnetoresistance in $R_{xx}$ is relatively small at high magnetic fields. As a result, the ordinary Hall term dominates the slope of $R_{xy}$ at high $B$, and the obtained value of $p$ is only weakly dependent on $n$ (see Fig. 2). As the magnetoresistance in $R_{xx}$ increases, the anomalous Hall term increasingly dominates the slope of the $R_{xy}$ versus $B$, and the uncertainty associated with the obtained value of $p$ increases.

The values of $p$ obtained are plotted against $T_C$ for the H-doped samples as well as for other (Ga,Mn)As films grown without H, in Fig. 3(a). The values of $T_C$ are obtained from the peak in the first derivative of resistivity versus temperature, $d\rho/dt$ [27]. Superconducting quantum interference device (SQUID) measurements of the remnant magnetic moment versus temperature yielded values of $T_C$ in good agreement with the ones obtained from the $d\rho/dt$ measurements, for both as-grown and fully annealed samples. For all samples studied, $T_C$ is found to increase monotonically as $p$ is increased by a succession of annealing steps. The same trend is observed for the electrical conductivity $\sigma$ (Fig. 3(b)), determined using standard 4-probe dc electrical measurements on the Hall bar structures.

To facilitate comparison with Ref. 11 as well as previous experimental and theoretical studies [6,7], the results are shown in a plot of $T_C/x_{eff}$ versus $p/N_{eff}$ in Fig. 4. $x_{eff}$ and $N_{eff} = 4x_{eff}/a$ represent the concentration of Mn magnetic moments which contribute to the magnetic order, and $a$ is the (Ga,Mn)As lattice constant [7]. $N_{eff}$ and $x_{eff}$ were estimated from SQUID magnetometry measurements of the low temperature magnetization, assuming a magnetic moment of $4\mu_B$ per $N_{eff}$, as expected for a S=5/2 local moment coupled antiferromagnetically to a valence band hole [28]. The concentrations of substitutional and interstitial Mn, $x_S$ and $x_I$, were then estimated by taking $x_{eff} = x_S$ in the fully annealed films, and $x_{eff} = (x_S - x_I)$ in the as-grown films, *ie* assuming an antiferromagnetic coupling between substitutional and interstitial moments [6,9,10,28]. Our analysis indicates that the compensation is very low in the fully annealed (Ga,Mn)As films, with approximately one hole per substitutional Mn ion.

Figure 4 shows that $T_C$ is not significantly suppressed in these weakly compensated, annealed (Ga,Mn)As thin films, in striking contrast to Ref. 11. The Mn and hole densities in Ref. 11 were obtained using ion channeling. Therefore, the lack of consistency with our high-field Hall results could potentially be ascribed to systematic uncertainties in our measurement and analysis procedure. However, we performed the same procedure for two samples where ion channeling results (obtained by a co-author of Ref. 11) have previously been reported [17], and found that the two techniques yield consistent values of $p$. The results are summarized in Table I, where it can be seen that the two methods are in agreement within the quoted uncertainty. The samples, which are 25nm thick films of annealed (Ga,Mn)As and (Al,Ga,Mn)As respectively, show lower than 10% compensation from both channeling and Hall measurements and a high $T_C$. These are indicated by the stars in Fig. 4.

Therefore, the lack of consistency between our results and those of Ref. 11 cannot be ascribed to the different measurement techniques, since these yield the same results when applied to the same high-quality samples. Instead, it most likely stems from qualitative differences in the defect structure of the (Ga,Mn)As samples studied by Dobrowolska *et al.* and of the (Ga,Mn)As films reported here. Ion channeling measures the concentrations of $Mn_S$ and $Mn_I$, but provides no information about other possible electrically active impurities. It has been shown previously [14] that growing (Ga,Mn)As with a high As/Ga ratio leads to a reduction of the $Mn_I$ concentration, accompanied by an increase in the concentration of As antisite ($As_{Ga}$) donors by approximately the same amount. This is consistent with the picture developed in Ref. 9 of a thermodynamic limit to the hole density in as-grown (Ga,Mn)As. It suggests that the samples described in Ref. 11 with low $T_C$, for which the hole density is inferred to be high in the as-grown state, may contain undetected compensating impurities such as $As_{Ga}$ which reduce $p$ from its inferred value. We also note that, in all the samples reported in Ref. 11, annealing leads to an increase in both $p/N_{eff}$ and $T_C/x_{eff}$, as well as a decrease in the low-temperature resistivity. This is hard to reconcile with the model put forward in Ref. 11 where $T_C$ is maximized around half-filling of an impurity band.

Figure 4 also compares the experimental data to the microscopic theory of Ref. 7. This is qualitatively consistent with our results at low compensation, since it predicts a high $T_C$ in this region rather than a strongly suppressed one. Differences are observed in the high compensation region, where the predicted suppression of ferromagnetism for $p/N_{eff} < 0.2$ is not observed. Samples in this region are highly resistive at low temperatures, and it is likely that localization of carriers plays an important role. For example, due to the short localization length it may be that the measured carrier density is that of delocalized holes, while the number of holes mediating the ferromagnetism is larger. Such layers, with low hole density and relatively high $T_C$, may be good candidates for showing a strong sensitivity of magnetic properties to electric fields [1].

In summary, we have conducted an investigation of the carrier density in a wide range of (Ga,Mn)As samples using the Hall effect at high magnetic fields and low temperatures. The results show that the carrier density in annealed (Ga,Mn)As films can be comparable to the substitutional Mn acceptor concentration, an observation which is supported by both SQUID magnetometry and ion channeling measurements. In these weakly compensated samples, there is no indication of the suppression of $T_C$ reported previously [11]. Our data demonstrate that the $T_C$ and electrical conductivity do not tend to zero at low compensation and thus the magnetic order in (Ga,Mn)As is *not* consistent with the picture of a Fermi level located in an isolated impurity band.

**Acknowledgements**
The work was supported by the UK EPSRC Grant No. EP/H002294/1, EU ERC Advanced Grant No. 268066, the Ministry of Education of the Czech Republic Grant No. LM2011026, and by the Praemium Academiae of the Academy of Sciences of the Czech Republic.

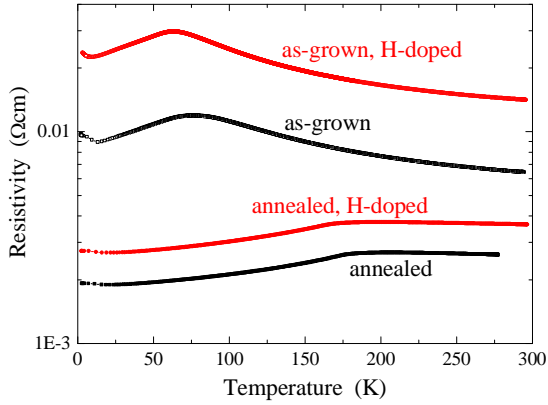

Fig. 1. Resistivity versus temperature for (Ga,Mn)As films with $x_{total}$=12%, as-grown and after annealing at 180ºC, with and without H co-doping.

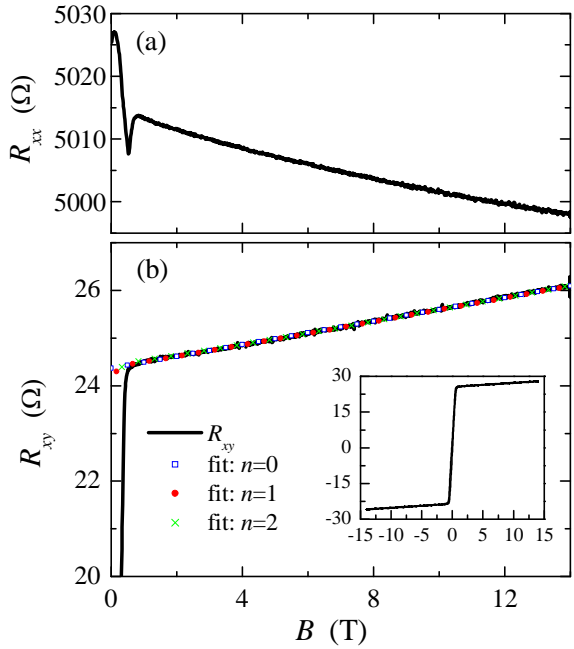

Fig. 2. (a) $R_{xx}$ and (b) $R_{xy}$ versus magnetic field $B$ measured at 2K, for a H-doped and annealed (Ga,Mn)As film with $x_{total}$=12%. The magnetic field is applied perpendicular to the film. Symbols in (b) are fits to $R_{xy}$ with exponent $n$=0 ($C$=24.4Ω), $n$=1 ($C$=0.0053) and $n$=2 ($C$=1.2×10$^{-6}$Ω$^{-1}$) respectively. The inset of (b) shows the full range of $R_{xy}$ on sweeping the magnetic field from negative to positive values.

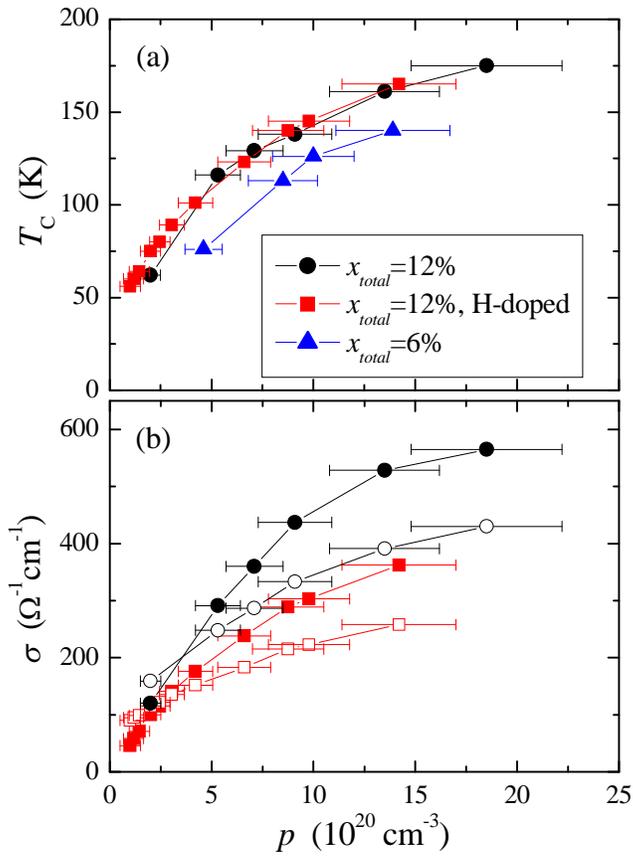

Fig. 3. (a) $T_C$ versus hole density obtained from high field Hall measurements, for 3 different (Ga,Mn)As films; (b) electrical conductivity at room temperature (open symbols) and low temperature (filled symbols) for (Ga,Mn)As films with $x_{total}$=12%, with (squares) and without (circles) H co-doping. In both (a) and (b), the hole density for each sample is varied by a series of short annealing steps.

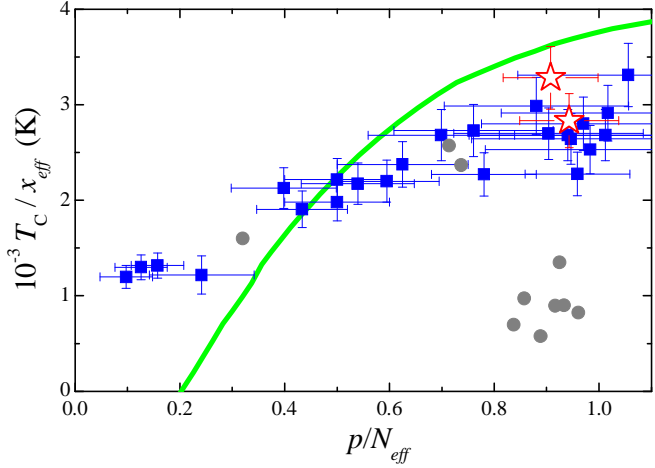

Fig. 4. $T_C/x_{eff}$ versus $p/N_{eff}$. Blue squares correspond to samples where hole density $p$ is obtained from high field Hall measurements. Gray circles correspond to samples from Ref. 11, where $p$ is obtained from ion channeling measurements. Red stars correspond to samples from Ref. 17, where $p$ is obtained from ion channeling measurements. The green line is the prediction of the microscopic calculation of Ref. 7.

| Sample | $T_C$ (K) | Ion channeling results | | | | $p$ from high field Hall ($10^{20}$ cm$^{-3}$) |
|---|---|---|---|---|---|---|
| | | $x_{sub}$ | $x_I$ | $x_{eff}$ | $p$ ($10^{20}$ cm$^{-3}$) | |
| $Ga_{0.94}Mn_{0.06}As$ | 128 | 0.043 | 0.004 | 0.039 | $7.8 \pm 0.9$ | $9.8 \pm 2.0$ |
| $Al_{0.1}Ga_{0.84}Mn_{0.06}As$ | 119 | 0.044 | 0.002 | 0.042 | $8.7 \pm 0.9$ | $8.5 \pm 1.7$ |

Table I. Comparison of ion channelling and Hall effect results: Curie temperatures ($T_C$), substitutional, interstitial and effective Mn concentrations ($x_{sub}$, $x_I$ and $x_{eff} = x_{sub} - x_I$) and hole concentration ($p = 4(x_{sub} - 2x_I)/a^3$) estimated by ion channeling, and hole concentration from high field Hall measurements, for annealed (Ga,Mn)As and (Al,Ga,Mn)As films. The ion channeling measurements are from Ref. 17.

---

[1] H. Ohno, D. Chiba, F. Matsukura, T. Omiya, E. Abe, T. Dietl, Y. Ohno, and K. Ohtani, Nature **408**, 944 (2000); A.W. Rushforth, E. De Ranieri, J. Zemen, J. Wunderlich, K.W. Edmonds, C.S. King, E. Ahmad, R.P. Campion, C.T. Foxon, B.L. Gallagher, K. Vyborny, J. Kucera, and T. Jungwirth, Phys. Rev. B **78**,